# An acoustically-driven biochip – Impact of flow on the cell-association of targeted drug carriers

Christian Fillafer[a], Gerda Ratzinger[a], Jürgen Neumann[b], Zeno Guttenberg[c], Silke Dissauer[a], Irene K. Lichtscheidl[d], Michael Wirth[a], Franz Gabor[a]*, Matthias F. Schneider[b]

[a]Department of Pharmaceutical Technology and Biopharmaceutics, Faculty of Life Sciences, University of Vienna, A-1090 Vienna, Austria. Fax: (+43)-1-4277-9554; Phone: (+43)-1-4277-55406; E-mail: franz.gabor@univie.ac.at

[b]University of Augsburg Experimentalphysik I – Biological Physics Group, 86135 Augsburg, Germany. Fax: (+49)-821-598-3225; Phone: (+49)-821-598-3311; E-mail: matthias.schneider@physik.uni-augsburg.de

[c]Olympus Life Science Research Europa GmbH, 81377 Munich, Germany.

[d]Cell Imaging and Ultrastructure Research, Faculty of Life Sciences, University of Vienna, A-1090 Vienna, Austria.

## Graphical contents entry

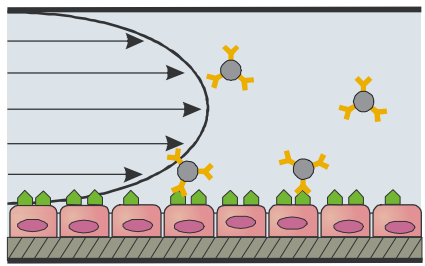

The accumulation of targeted drug carriers in the diseased tissue is substantially affected by the locally acting hydrodynamic drag forces. A thumbnail-sized microfluidic chip incorporating a surface acoustic wave pump can be used to study the adhesion of particles to cells under physiological flow conditions.

## Summary

The interaction of targeted drug carriers with epithelial and endothelial barriers *in vivo* is largely determined by the dynamics of the body fluids. To simulate these conditions in binding assays, a fully biocompatible *in vitro* model was developed which can accurately mimic a wide range of physiological flow conditions on a thumbnail-format cell-chip. This acoustically-driven microfluidic system was used to study the interaction characteristics of protein-coated particles with cells. Poly(D,L-lactide-co-glycolide) (PLGA) microparticles (2.86 ± 0.95 µm) were conjugated with wheat germ agglutinin (WGA-MP, cytoadhesive protein) or bovine serum albumin (BSA-MP, nonspecific protein) and their binding to epithelial cell monolayers was investigated under stationary and flow conditions. While mean numbers of 1500 ± 307 mm$^{-2}$ WGA-MP and 94 ± 64 mm$^{-2}$ BSA-MP respectively were detected to be cell-bound in the stationary setup, incubation at increasing flow velocities increasingly antagonized the attachment of both types of surface-modified particles. However, while binding of BSA-MP was totally inhibited by flow, grafting with WGA resulted in a pronounced anchoring effect. This was indicated by a mean number of 747 ± 241 mm$^{-2}$ and 104 ± 44 mm$^{-2}$ attached particles at shear rates of 0.2 s$^{-1}$ and 1 s$^{-1}$ respectively. Due to the compactness of the fluidic chip which favours parallelization, this setup represents a highly promising approach towards a screening platform for the performance of drug delivery vehicles under physiological flow conditions. In this regard, the flow-chip is expected to provide substantial information for the successful design and development of targeted micro- and nanoparticulate drug carrier systems.

## Introduction

Physiological and pathological processes such as the site-specific adhesion of platelets, leukocytes or metastasizing cancer cells underlie sophisticated mechanisms in order to efficiently function in the presence of hydrodynamic flow. Substantial knowledge about these processes has been generated by simulating physiological shear conditions with *in vitro* fluidic systems such as the parallel plate flow chamber (PPFC) and the radial flow detachment assay (RFDA).[1-4] Although primarily aimed at understanding physiology, these fundamental studies also bear essential implications for the development of targeted colloidal drug carriers.[1, 3, 5] Presently, the target effect of site-specific drug delivery systems is by default determined with *in vitro* cell binding assays under stationary conditions. However, regarding the extent and specificity of particle binding, recent reports have identified clear discrepancies between the results obtained from stationary and more realistic, dynamic models of the *in vivo* environment.[6-8] Particularly, the presence of substantial hydrodynamic drag forces upon application *in vivo* is expected to explicitly affect the deposition characteristics of ligand-coated particles.[7, 9, 10] The peristaltic motion in the gastrointestinal tract, for example, leads to streaming velocities of ~85 μm s$^{-1}$ in the jejunum and ~55 μm s$^{-1}$ in the ileum.[11] The variation of flow conditions in the circulatory system is even more pronounced as illustrated by effective shear rates ranging from 1 s$^{-1}$ in wide vessels to 10$^5$ s$^{-1}$ in small arteries.[12] To attain preferential binding of the carrier to the diseased tissue in this environment, the size and ligand coating density of the colloids has to be adjusted according to the flow conditions as well as the expected receptor density and affinity at the target tissue.[1, 2, 5] In order to be able to practically optimize these parameters of potential drug delivery vehicles, parallelisable *in vitro* bioassays have to be developed which offer the possibility of controllable flow generation. At this, increased experimental throughput is highly necessary in order to cope with the extensive amount of samples which have to be processed to generate sufficient data sets. Although well suited for specific questions, the PPFC and RFDA are of limited use for such applications. When multiple experiments are demanded at high reproducibility, the complexity of these methods' tubing and chambers becomes cumbersome to handle. Microfluidic pumping techniques might allow for approaching these shortcomings successfully. The most common methods to generate flow in miniaturised systems are thermophoresis (via Marangoni forces), electrochemical reactions, surface patterning methods, electro-osmosis, induced charge electro-osmosis as well as several mechanical pumping strategies.[13, 14] However, these techniques' suitability is limited for systems which are required to comply with cell growth and cell viability. For such applications a micropumping technology based on acoustic streaming might represent a highly promising alternative. This technique generates locally defined surface acoustic waves (SAWs) which, when coupled into liquid, result in a defined pressure gradient that can be utilized for streaming (Figure 1A).

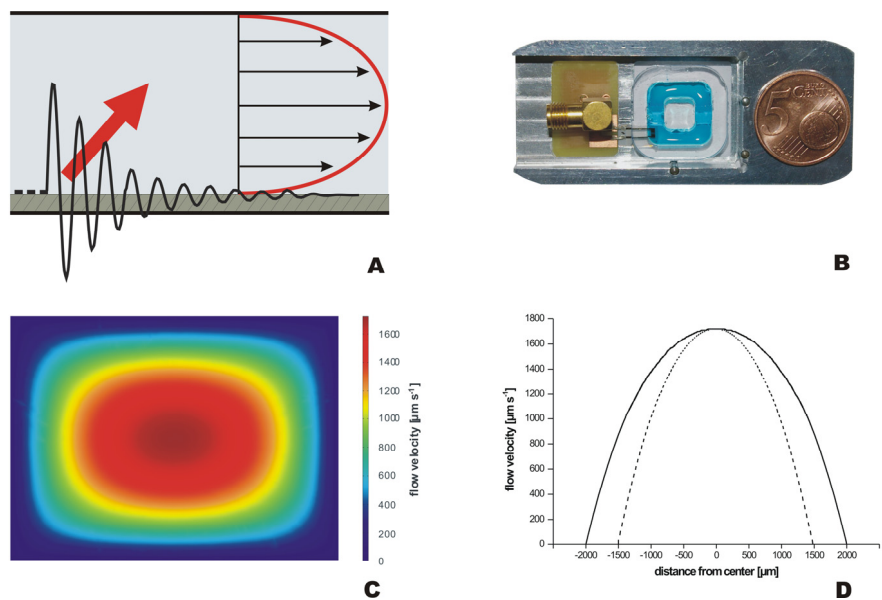

**Figure 1.** SAWs are generated on a piezoelectric chip and lead to fluid streaming with a parabolic flow profile when coupled into a liquid-filled channel (A). Dimensions of the SAW-chip with a positioned 3D-microchannel (B). Cross section of the flow velocity profile generated in a 3D-microchannel by acoustic streaming 4 mm x 3 mm (width x height) (C). Horizontal (D, solid line) and vertical cut (D, dashed line) through (C).

In the work presented, the applicability of this technology to controllably produce fluid flow in a miniaturized pharmaceutically relevant *in vitro* bioassay was tested. At this, the effect of shearing on the association of protein-decorated particles with epithelial cell monolayers was investigated. Biocompatible poly(D,L-lactide-co-glycolide) (PLGA) microparticles (MP) conjugated with fluorescence labelled wheat germ agglutinin (WGA) and bovine serum albumin (BSA) served as representative targeted and non-targeted model drug carrier systems. This model system was chosen, since studies have shown that decoration with WGA mediates binding to Caco-2 cells under stationary conditions.[15, 16] The extent of a targeting effect in the presence of hydrodynamic drag was investigated using the micropumping SAW-chip (Figure 1B) to simulate physiological flow conditions.

## Results and discussion

### The acoustically-driven microfluidic system

Similar to conventional flow assays such as the PPFC and RFDA, the acoustically-driven flow chip is based on a fluidic channel which is directly accessible during the experiment via microscopy. However, in contrast to the previously mentioned assays that are driven by external mechanical pumps, the developed microfluidic chip uses a planar, non invasive pumping principle for flow generation. This technique realizes the contamination-free pumping of liquid volumes as small as a few microliters by surface acoustic streaming which can generate shear rates between 0.01 $s^{-1}$ and 1000 $s^{-1}$ depending on the system's geometry. As a consequence of the high frequency signal based generation of the SAW, the pumping performance is continuously actuated and varied via the parameters of a conventional signal generator (see Video S1 of the Supporting Information).[17] By channelizing the streamed liquid into 3D-microchannels, which are readily fabricated in almost arbitrary geometries by elastomeric molding, a flow chamber can be created. In this study, a rather simple 44 mm x 4 mm x 3 mm (length x width x height) racetrack-format poly(dimethylsiloxane) (PDMS)-cast was used (Figure 1B). The channel was dimensioned in such a manner that sufficient medium could be included for uncomplicated cell culturing of epithelial Caco-2 monolayers. Thus, the usually necessary transfer of a cell-covered substrate to the flow chamber or alignment of an elastomeric cast on the cultured cells can be avoided. However, the pumping technique is not limited to rectangular casts of this size, but can be easily adapted to drive flow in smaller structures with constrictions or bifurcations.[17] Channels including the latter geometries could be exceptionally insightful tools for a realistic simulation of the complex vessel-flow in the circulatory system.[9] At this, it is additionally helpful, that not only continuous flow modes can be accurately mimicked on the SAW-chip, but that also precisely defined pulsed pumping is possible. While being hardly achievable with mechanical pumps, this streaming mode is controllably induced by amplitude modulation of the high frequency signal (see Video S2 of the Supporting Information). Thereby, the effects of pulsating flow combined with defined shear rates could be collectively investigated regarding their impact on distribution and adhesion processes in the branched circulatory system.[9]

Besides its compliance with microfluidic channels, which allows for reducing the necessary amount of reagents to a minimum, the chip-based setup additionally bears the advantage of omitting tubing and connectors which are otherwise necessary to pipeline liquid within the system. Consequently, the dead volume associated with mechanical pumps is avoidable and the entire device can be downsized notably (Figure 1B). This feature is essential, as it permits parallelization of several SAW-chips in one platform thus increasing the amount of samples which can be processed simultaneously. Even a setup with multiple SAW-pumps and channels in a microplate format is realizable, entailing the ability to use microplate readers for analysis. Such an extension to more sensitive analytical methods could be a powerful tool to gain information on adhesion phenomena involving nanoparticles or proteins, since analyses based on microscopic imaging hit on their limits in this size regime.

**Cell-binding of surface-modified microparticles under stationary conditions**

To establish reference values for the experiments involving fluid flow, the extent of binding of BSA- and WGA-MP to Caco-2 monolayers was primarily determined in a stationary setup (flow velocity$_{max}$ = 0 µm s$^{-1}$). For the used microparticles with a mean size of 2.86 ± 0.95 µm and a density of ~ 1.28 g cm$^{-3}$ a sedimentation controlled particle deposition is expected.[18] In the equilibrium state between frictional force and gravitation, settling of the colloids occurs with a constant z-velocity of 1.4 µm s$^{-1}$. Consequently, 83% of the homogenously distributed particles are assumed to deposit on the surface within 30 min leading to a maximum coverage of 3500 mm$^{-2}$. In practice, however, only 94 ± 64 mm$^{-2}$ BSA-MP were detected after stationary loading (30 min), washing and stationary chase-incubation (30 min) (Figure 2, stars). The rather low number of cell-associated particles, which corresponds to 3% of the theoretical load, can be explained by the reportedly low affinity of albumin-coated PLGA-particles to Caco-2 cells.[6, 15] Due to the lack of interactive strength the washing steps resulted in the detachment of all except a few non-specifically bound particles. In contrast, conjugation with f-WGA led to a clearly enhanced adhesion of colloids to the monolayer. This is in line with previous studies which have identified WGA as an agent for enhancing adhesion to Caco-2 cells.[19, 20] Following incubation under stationary conditions a mean number of 1500 ± 307 mm$^{-2}$ cell-associated particles, which corresponds to 43% of the maximum load, were detected. This 16-fold increased interaction as compared to BSA-MP is explicitly illustrated in Figure 3 and can be attributed to specific binding of the lectin to membrane-associated N-acetyl-D-glucosamine and N-acetyl-neuraminic-acid residues.[15, 21] WGA-MP which had not bound to the cell membrane were removed in course of the washing steps.

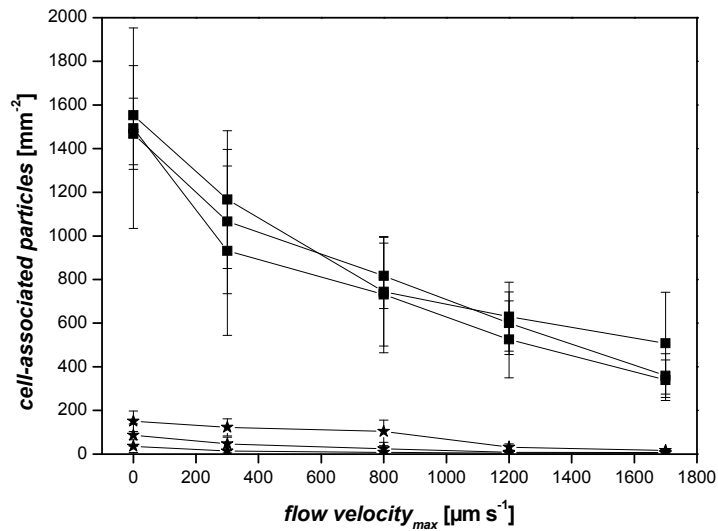

**Figure 2.** Mean number of WGA- (squares) and BSA-MP (stars) associated with Caco-2 monolayers after loading for 30 min at stationary conditions, washing and chase-incubation under stationary or flow conditions. Each set of data points was obtained from independent experimental series.

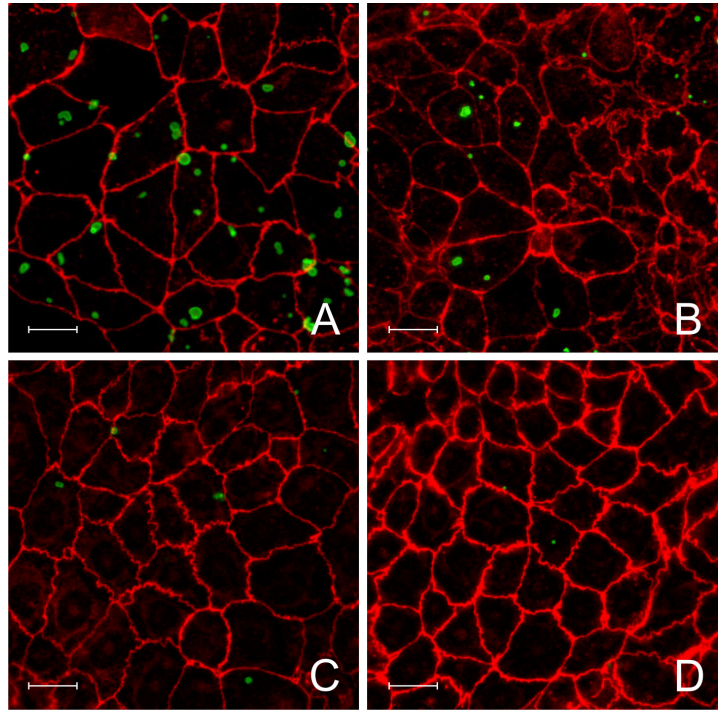

**Figure 3.** Cell-associated WGA-MP (A) and BSA-MP (C) respectively after loading for 30 min, washing and chase-incubation under stationary conditions (flow velocity$_{max}$ = 0 μm s$^{-1}$). Chase-incubation performed under flow conditions (flow velocity$_{max}$ = 1700 μm s$^{-1}$; B and D respectively). Microparticles (green) and tight junction associated protein ZO-1 (red). Bar represents 20 μm.

**Impact of flow on cell-bound particles**

To investigate the effect of different shear rates on cell-associated BSA- and WGA-MP, stationary particle-loaded Caco-2 monolayers were chase-incubated under flow conditions. At this, acoustic streaming induces laminar fluid flow in the channel and thereby generates hydrodynamic forces acting on the attached microparticles. If the adhesive bonds break and do not re-establish elsewhere, the fluid flow transports the colloids to the coupling region of the SAW, where the considerable lift forces redisperse them in the channel cross-section. Consequently, most of these particles are not available for reattachment to the cells. Using this setup, a shear rate dependent reduction of the number of cell-bound particles was monitored for both types of surface-modified colloids (Figure 2). While low shear rates (flow velocity$_{max}$ = 300 µm s$^{-1}$ and 800 µm s$^{-1}$) led to the detachment of ~30% and 50% respectively of the initially cell-associated BSA-MP, incubation at higher shear rates (flow velocity$_{max}$ = 1200 µm s$^{-1}$ and 1700 µm s$^{-1}$) almost completely removed the albumin-conjugated colloids from the cell-surface (Figure 3D). Obviously, the few BSA-MP, which were associated with the Caco-2 monolayer after stationary loading and washing, are characterized by low adhesivity which is not sufficient to anchor the particles in the presence of shear forces. In contrast, conjugation of microparticles with carbohydrate-binding protein not only led to higher cell-binding under stationary conditions but also enhanced retention on the Caco-2 monolayer in the presence of flow (Figure 2, squares). This is illustrated by a mean number of 1057 ± 351 mm$^{-2}$ and 400 ± 168 mm$^{-2}$ monolayer-associated colloids at the lowest and highest flow velocity studied. Consequently, as compared with BSA-MP, WGA-MP were characterized by at least 17-fold increased retention over the whole range of flow velocities investigated. Interestingly, in case of the highest shear rates, the effect of the lectin-corona was even more pronounced as exemplified by 39-fold and 44-fold improved adhesion over albumin-conjugated colloids (Figure 3C).

Regarding reproducibility of the binding assays on the SAW-chip, it should be highlighted that the three data series for each particle type plotted in Figure 2 were obtained from separate experiments. The low deviation between the curves underlines the SAW-pump's ability to controllably and reproducibly generate flow in the 3D-microchannels, which is a crucial prerequisite that determines the practical usability of the system. The standard deviation of each data point is comparable with those in similar studies and very likely a result of the image-based quantification combined with the cellular substrate inherent characteristics.[7] To minimize this, homogenous receptor-coated surfaces could be used, albeit at the cost of setting aside the complexly constituted cell membrane. Hence, when probing the interaction of targeted drug carriers with biological barriers, a substitution of the cell monolayer by an artificial substrate is not expedient since it reduces the relevance for comparisons with the *in vivo* conditions.[3, 7, 8] When studying isolated adhesion phenomena, however, the incorporation of precisely surface-engineered substrates is certainly realisable as well as preferable in terms of analytical accuracy.[5, 12, 22]

**Cell-binding of surface-modified microparticles under flow conditions**

**Impact of flow on particle deposition in the channel.**

To estimate the effect of flow on the deposition rate of particles in the 3D-microchannels, the stationary condition was compared with streaming. The experimental microfluidic setup is characterized by Reynolds numbers of Re = 1 and Re = 6 for the lowest and highest velocity respectively which indicates laminar flow. As determined by computational fluid dynamics simulations the flow profile in the channel is parabolic (Figure 1C, D). Considering these conditions and that the images used for quantification of the adherent particles were taken in a small central region of the channel, the lateral y-component of the velocity was assumed to be widely independent from the vertical z-component. Therefore, a particle's trajectory in the flowing fluid contains components in the direction of flow (x) as well as in the vertical direction (z) with the former one depending on the latter. At the start of the experiment, the 3D-microchannels are filled with buffer containing $7.5 \times 10^5$ homogenously distributed microparticles which sediment with a constant z-velocity. Upon engaging the flow, these particles are travelling for one channel length until again reaching the coupling zone of the SAW-pump where they are homogenously redistributed over the cross-section of the channel. On each of these rotations, particles below a critical height $\Delta z_{crit}$ settle out on the cell monolayer. For a flow velocity $v_{max}$ of 1700 µm s$^{-1}$ the critical height $\Delta z_{crit}$ is determined to be ~224 µm. Taking into account that the particle density in the channel decreases with duration of the experiment, ~56% of the microparticles are expected to have deposited on the cell monolayer after 30 min. This corresponds to a theoretical surface coverage of 2400 mm$^{-2}$.

In the case of lower flow velocities the increased critical height $\Delta z_{crit}$ interestingly does not lead to a higher deposition rate since its effect is widely levelled out by the proportionally lowered particle flux density. Consequently, almost equal particle deposition rates on the cell monolayer were to be expected under all investigated flow conditions.

**Binding of BSA- and WGA-MP to a Caco-2 monolayer under flow conditions.**

In order to investigate the binding of BSA- and WGA-MP from a streaming medium to Caco-2 monolayers, the particle suspension was transferred to the 3D-microchannels and fluid flow was instantly engaged. As illustrated in Figure 4 (open symbols), incubation under flow conditions led to a clearly decreased cell binding of the ligand-conjugated colloids. Partially, this effect is system inherent and can be attributed to the previously discussed reduction of particle deposition due to the parabolic flow in the channel. Taking this into account and considering the binding potential of 3% and 43% as determined in the absence of hydrodynamic forces, a maximal surface coverage of ~70 mm$^{-2}$ and ~1000 mm$^{-2}$ is expected for BSA- and WGA-MP respectively under flow conditions. This estimate seems quite realistic as illustrated by 77 ± 31 mm$^{-2}$ and 747 ± 241 mm$^{-2}$ respectively bound particles in case of the lowest shear force. However, aside from the generally lowered base deposition, the number of cell-bound colloids was additionally diminished by the antagonizing effect of flow on particle-surface bonds.[4, 5] For BSA-MP this increased hydrodynamic drag resulted in a nearly complete inhibition of the association with the Caco-2 monolayer. This is explicated by a marginal surface coverage of 34 ± 16 mm$^{-2}$ cell-associated particles at the highest shear rate. Apparently, albumin-modified particles which are deposited on the monolayer under flow conditions do not notably interact with the cell membrane. Similar conclusions have been described in a recent study where a gas lift driven ussing chamber setup was used to simulate the effect of gastrointestinal flow on particle binding.[6] In contrast to albumin-conjugated colloids, WGA-MP exhibited higher binding at all investigated flow-conditions. The advantage of lectin-conjugation was clearest at low shear rates. However, even at the highest flow velocities studied at least threefold more particles associated with the cell monolayer as compared with BSA-MP. These observations lead to the conclusion that colloids lacking appropriate surface chemistry do not notably attach to Caco-2 cells at relatively moderate shear rates ranging from 0.2 s$^{-1}$ to 1 s$^{-1}$. When considering this combined with the very low cytoadhesion under stationary conditions, the use of BSA-MP as drug carriers is limited. The use of plain colloids for most pharmaceutically relevant applications has to be relativized even more due to the reportedly marginal cytoadhesion of unconjugated PLGA particles.[15] To efficiently compliment the advantageous biocompatibility and biodegradability of particles made from PLGA and similar polymers, surface modification with targeting ligands is essential. In this regard, however based on stationary studies, wheat germ agglutinin has been proposed as a promising candidate to serve this purpose for peroral applications. Due to its carbohydrate binding properties, which mediate adhesion to enterocytes as well as mucus, a prolonged gastrointestinal residence time of sustained release drug carriers might be achieved.[6, 15, 23] Using the SAW-driven microfluidic device, it was possible to show that the interaction between WGA and the Caco-2 cell's glycocalyx is indeed sufficient to mediate the binding of 3µm-sized particles under physiological flow conditions. This observation further underlines the potential of wheat germ agglutinin for peroral delivery, where low shear rates act on the carrier. However, in the presence of higher flow velocities, the studied WGA-MP exhibit a propensity to detach from the cell layer. This could be counteracted by using smaller particles, possibly in the nanometer size range, since the effective hydrodynamic drag forces decrease with particle diameter.[1] Moreover, conjugation of particles with alternative targeting moieties could lead to enhanced cytoadhesion under the shear rates encountered in the circulatory system. In this regard, the underlying mechanisms of shear-activated proteinic ligands like the FimH subunit of Type-1 fimbriae of *E. coli* or von Willebrand factor might lead the way for the engineering of site-specific drug carriers that efficiently adhere under flow conditions.[12, 24]

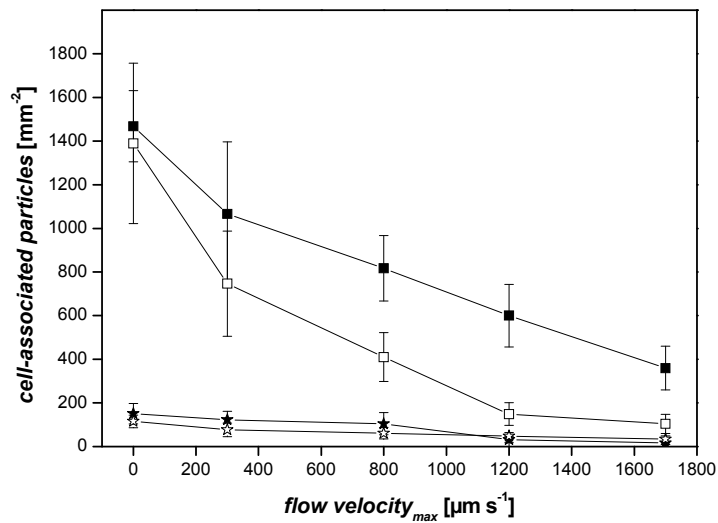

**Figure 4.** Mean number of WGA-MP (squares) and BSA-MP (stars) associated with Caco-2 monolayers upon incubation under stationary and flow conditions. Monolayers pre-loaded with microparticles for 30 min under stationary conditions (filled symbols). Direct incubation of microparticles with monolayers under stationary and flow conditions (open symbols).

## Experimental

**Materials**

Sylgard® 184 Silicone Elastomer Kit was purchased from Baltres (Baden, Austria). Resomer® RG502H (PLGA, lactide/glycolide ratio 50:50, inherent viscosity 0.22 dL g$^{-1}$, acid number 9 mg KOH g$^{-1}$) was obtained from Boehringer Ingelheim (Ingelheim, Germany). Fluorescein-labeled wheat germ agglutinin (molar ratio fluorescein/protein (F/P) = 2.9) from *Triticum vulgare* was bought from Vector laboratories (Burlingame, USA). FITC-labeled bovine serum albumin (F/P = 12), N-(3-Dimethylaminopropyl)-N′-ethylcarbodiimide hydrochloride (EDAC), N-hydroxysuccinimide (NHS), and Pluronic® F-68 were purchased from Sigma Aldrich (Vienna, Austria). All other chemicals used were of analytical purity.

**Preparation of functionalized microparticles**

PLGA microparticles with a mean diameter of 2.86 ± 0.95 μm were prepared by spray drying of a 6.5% (w/v) solution of PLGA in dichloromethane with a Buechi Mini Spray Dryer B-191 (Buechi, Flawil, Switzerland) as previously described.[19] For surface modification, 100 mg of the PLGA microparticles were suspended in 20 mM HEPES/NaOH pH 7.0 (10 mL) and activated with solutions of EDAC (360 mg in 1.5 mL) and NHS (15 mg in 1 mL) in the same buffer for 2 h under end-over-end rotation at room temperature. In order to remove excess coupling reagent, the suspension was diluted threefold with 20 mM HEPES/NaOH pH 7.4 and centrifuged (10 min, 2500 rpm, 4°C). The resulting pellet was resuspended in 20 mM HEPES/NaOH pH 7.4 (10 mL). Upon addition of F-WGA (1.00 mg) and F-BSA (1.83 mg) respectively, end-over-end incubation was performed overnight at room temperature. Remaining active ester intermediates were saturated by addition of glycine (450 mg) in 20 mM HEPES/NaOH pH 7.4 (6 mL) and further incubation for 30 min. Subsequently, the microparticles were washed three times by centrifugation (10 min, 3200 rpm, 4°C) and resuspension in 20 mM HEPES/NaOH pH 7.4 (30 mL). After the last centrifugation step, the particles were suspended in a solution of 0.1% Pluronic F-68 in isotonic 20 mM HEPES/NaOH pH 7.4 (10 mL).

**Fabrication of sterile 3D-microchannels**

Base (10 g) and curing agent (1 g) of the silicone elastomer kit were mixed in a test tube, vigorously stirred and evacuated for 30min to remove gas bubbles. After pouring the liquid prepolymer into pre-structured aluminium molds and hardening over night at 70°C, the PDMS replicas were peeled from the master and placed on 24 x 24 mm (length x width) glass cover slips. Following assembly, the 3D-microchannels dimensioned 44 x 4 x 3 mm (length x width x height) were transferred to glass Petri dishes and autoclaved for 50 min at 121°C (1 bar).

**Cell Culture in 3D-microchannels**

The Caco-2 cell line was purchased from the German collection of microorganisms and cell culture (DSMZ, Braunschweig, Germany). Tissue culture reagents were obtained from Sigma (St. Louis, USA) and Gibco Life Technologies Ltd. (Invitrogen Corp., Carlsbad, USA). Cells were cultivated in RPMI 1640 cell-culture medium containing 10% fetal calf serum, 4 mM L-glutamine and 150 μg mL$^{-1}$ gentamycine in a humidified 5% $CO_2$/95% air atmosphere at 37°C and subcultured with Tryple Select from Gibco (Lofer, Austria). For the microfluidic experiments, each sterile 3D-microchannel was filled with 500 μL of Caco-2 single cell suspension (1.36 x 10$^5$ cells mL$^{-1}$) and cultivated under standard cell culture conditions until a confluent cell monolayer had formed.

**Preparation of SAW-chip**

LiNbO$_3$ slides (128°-cut x-propagation) dimensioned 15 x 15 x 0.4 mm (length x width x height) were used as piezoelectric substrates. Interdigital metal structures (IDTs) were structured on these slides by standard lithographic processes in order to predominately generate (Rayleigh-mode) SAWs.[25] The used IDTs had 42 fingerpairs, an aperture of 600 μm and a periodicity of 26 μm, resulting in a resonance frequency of about 153 MHz. To enhance the resistance against mechanical cleaning procedures, the fingers were additionally coated with a radio frequency (RF)-sputtered SiO$_2$ protective coating.

**Microparticle-cell interaction studies at stationary and flow conditions**

For the interaction studies the BSA-MP and WGA-MP were suspended in isotonic 20 mM HEPES/NaOH pH 7.4 at a concentration of 7.5 x 10$^5$ particles per 500 μL. Shortly before the experiment, the cell culture medium was removed from the 3D-microchannels and the monolayers were washed once with isotonic 20 mM HEPES/NaOH pH 7.4 (500 μL). Subsequently, the microparticle suspension was added and the channels were covered with a glass cover slip. In order to grant efficient transmission of the SAWs, 50 μL of water were pipetted on the piezoelectric chip as a coupling fluid before the 3D-microchannel was placed on top of it. This setup was mounted on a fluorescence microscope and connected to the high frequency generator, which had been configured to supply preset energy inputs corresponding to flow velocities of 0 μm s$^{-1}$, 300 μm s$^{-1}$, 800 μm s$^{-1}$, 1200 μm s$^{-1}$ and 1700 μm s$^{-1}$ respectively. Following incubation at either stationary or flow conditions for 30 min, the monolayers were washed three times with isotonic 20 mM HEPES/NaOH pH 7.4 (500 μL) and embedded as described below. For an alternative set of experiments, monolayers loaded for 30 min under stationary conditions were washed and subsequently subjected to stationary or flow chase-incubation. After two additional washing steps with isotonic 20 mM HEPES/NaOH (500 μL) the PDMS-structure was peeled off these channels as well. The glass cover slips with the adherent monolayers were embedded in a drop of FluorSave$^{TM}$ (Calbiochem$^{®}$, USA and Canada) and were stored at 4°C for 12 hours prior to further analyses.

**Fluorescence microscopy and software-based analysis**

The embedded Caco-2 monolayers were analysed on a Nikon Eclipse 50i microscope (Nikon Corp., Japan) equipped with an EXFO X-Cite 120 fluorescence illumination system. A random series of non-overlapping fluorescence microscopic images (n = 6) was acquired over the channel area, whereby care was taken to analyze monolayer parts located in the centre of the 3D-microchannels. To grant comparability, the settings of the fluorescence lamp and exposure time were left constant during the data acquisition process. Finally, the number of cell-associated microparticles in every image was determined with the threshold-dependant automated particle analysis of ImageJ (NIH, USA). The number of cell-associated particles represents the mean value which was calculated from the images acquired in independent channels (n = 2).

**Antibody staining**

After incubation at stationary and flow conditions respectively, the monolayers were fixed with a 2% solution of paraformaldehyde for 15 min at room temperature and were washed with phosphate buffered saline pH 7.4 (PBS; 500 μL). Upon treatment with a 50 mM solution of NH$_4$Cl for 15 min and with a 0.1% solution of Triton X-100 for 10 min, the cells were washed again with PBS (500 μL). The tight junction associated protein ZO-1 was stained for 1 h at 37°C with a primary antibody (BD Biosciences, San Jose, USA) diluted 1:100 in a 1% solution of BSA in PBS. Upon washing thrice with a 1% solution of BSA in PBS, the monolayers were incubated with a 1:100 dilution of a secondary Anti-Mouse Immunoglobulin-RPE antibody (Dako Denmark A/S, Glostrup, Denmark) for 30 min at 37°C. Finally, the cell layer was washed three times with a 1% solution of BSA in PBS and mounted in a drop of FluorSave$^{TM}$.

## Conclusion

Stationary binding assays have become the current standard in preclinical biopharmaceutical testing due to the rather simple handling and lack of alternative *in vitro* models. To approach this shortcoming, an acoustically-driven thumbnail-sized microfluidic chip was developed that can controllably and reproducibly generate flow in 3D-microchannels which are compatible with cell culture. This SAW-chip was used to investigate the binding properties of albumin- and wheat germ agglutinin-conjugated microparticles to an epithelial cell layer under flow conditions. As illustrated in the present work, the results obtained from binding assays under stationary conditions notably differ from those obtained in systems simulating the dynamic *in vivo* environment. It was found that non-targeted microparticles possess a very low propensity to bind and are detached in the presence of flow, while conjugation with WGA led to a distinctly improved adhesion of particles to the cell-layer at shear rates ranging from 0.2 s$^{-1}$ to 1 s$^{-1}$. In conclusion, these results clearly underline the importance of surface functionalization for the design of nano- and microparticulate drug delivery vehicles. Elaborate models of the *in vivo* conditions are necessary in order to realistically predict and optimize the performance of such systems prior to animal studies. In this context, the developed microfluidic SAW-chip is expected to provide a highly versatile platform for an investigation of flow-associated effects on particle-cell adhesion processes. Fundamental information distilled from studies using this technology will deeply benefit the engineering of artificial drug carrier systems which perform with high efficiency in the dynamically complex physiological environment.

## Acknowledgements

The authors thank U. Länger and Y.X. Wang for help with preparation of the microspheres, Rohde&Schwarz GmbH for assistance with the high frequency generator and K. Sritharan, S. Nuschele for helpful discussions. Parts of this work were supported by the CellPROM project, funded by the European Community as contract No. NMP4-CT-2004-500039 under the 6[th] Framework Programme for Research and Technological Development in the thematic area of "Nanotechnologies and nanosciences, knowledge-based multifunctional materials and new production processes and devices". The contribution reflects the author's views and the community is not liable for any use that may be made of the information contained therein.